\documentclass{article}
\usepackage{spconf,amsmath,graphicx}

\usepackage{booktabs}
\usepackage{amsmath}
\usepackage{enumitem}
\usepackage{graphicx}
\usepackage{xcolor} 
\usepackage{tabularx} 
\setlist{nosep, leftmargin=14pt}
\usepackage{pifont}%
\usepackage{mwe} 
 \usepackage{url}
 \usepackage{multirow}
 
\title{Liver Cirrhosis Stage Estimation from MRI with Deep Learning}

\name{\parbox{\linewidth}{\centering Jun Zeng$^{1}$, Debesh Jha$^{3}$, Ertugrul Aktas$^{2}$, Elif Keles$^{2}$,  Alpay Medetalibeyoglu$^{2}$, Matthew Antalek$^{2}$, \\Federica Proietto Salanitri$^{2}$, Amir A. Borhani$^{2}$, Daniela P. Ladner$^{2}$, Gorkem Durak$^{2}$, Ulas Bagci$^{2}$}\thanks{The project is supported by the NIH funding: R01-CA246704, R01-CA240639, U01 DK127384-02S1, and U01-CA268808.}}
 
\address{$^{1}$Chongqing University of Posts and Telecommunications, Chongqing, China \\
	$^{2}$Machine and Hybrid Intelligence Lab, Northwestern University, Chicago, IL, USA. \\
	$^{3}$University of South Dakota, Vermilion, SD, USA 
} 

\begin{document}

\maketitle

\begin{abstract}

We present an end-to-end deep learning framework for automated liver cirrhosis stage estimation from multi-sequence MRI. Cirrhosis is the severe scarring (fibrosis) of the liver and a common endpoint of various chronic liver diseases. Early diagnosis is vital to prevent complications such as decompensation and cancer, which significantly decreases life expectancy. However, diagnosing cirrhosis in its early stages is challenging, and patients often present with life-threatening complications. Our approach integrates multi-scale feature learning with sequence-specific attention mechanisms to capture subtle tissue variations across cirrhosis progression stages. Using \textit{CirrMRI600+}, a large-scale publicly available dataset of 628 high-resolution MRI scans from 339 patients, we demonstrate state-of-the-art performance in three-stage cirrhosis classification. Our best model achieves 72.8\% accuracy on T1W and 63.8\% on T2W sequences, significantly outperforming traditional radiomics-based approaches. Through extensive ablation studies, we show that our architecture effectively learns stage-specific imaging biomarkers. We establish new benchmark (spanning from VGGs to ResNets, and Mamba) for automated cirrhosis staging and provide insights for developing clinically applicable deep learning systems. The source code is available at \url{https://github.com/JunZengz/CirrhosisStage}.

\end{abstract}

\section{Introduction}
\label{sec:introduction}
Cirrhosis and its devastating complications present a significant global health challenge with rising mortality rates~\cite{wang2023remodeling,huang2023global}. Responsible for nearly one and a half million deaths annually worldwide, cirrhosis accounts for 2.4\% of all global fatalities~\cite{huang2023global,liu2019wjg}. Over the past two decades, mortality rates have increased by 26.4\%, disproportionately affecting younger populations aged 25 to 34 years~\cite{lee2024designing}. The year following a diagnosis of cirrhosis is when individuals are most vulnerable, highlighting the importance of early detection and intervention~\cite{daher2022proportion}. The most decisive prognostic factor for survival is the presence or absence of hepatic decompensation, as progression to decompensation significantly increases medical costs and mortality rates~\cite{sepanlou2020global,lee2023evolution}. Approximately 11\% of individuals with compensated cirrhosis experience new decompensations each year, and this progression is associated with a decline in median survival from 12 years to less than 2 years~\cite{lee2023evolution}.  Moreover, liver cancer ranks among the top six most common cancers worldwide and the third most common cause of cancer death~\cite{sung2021global}. HCC accounts for approximately 80\% of liver cancer cases, with 80\%-90\% of individuals with HCC having underlying cirrhosis~\cite{yang2019global,fattovich2004hepatocellular}. Therefore, early detection and accurate staging of cirrhosis are crucial for preventing complications and ensuring effective disease management and treatment planning.

Cirrhosis is characterized by the gradual replacement of healthy liver tissue with irreversible scarring or fibrosis due to various chronic liver diseases~\cite{ullah2023biological}.  The liver shows no significant morphological changes during the early stages of cirrhosis. As a result, diagnosing cirrhosis at an early stage is difficult and is often missed, even by expert specialists. If the diagnosis is missed in the early stages and the disease progresses, it results in various complications that severely affect patient outcomes~\cite{premkumar2022overview}. Traditional diagnosis methods rely on clinical symptoms, laboratory tests, and ultrasound imaging, which restrict their ability to identify subtle tissue changes. In contrast, MRI provides higher image quality and improved accuracy for diagnosing cirrhosis and assessing its stages, allowing physicians to evaluate the liver’s structure and tissue changes with precision.

As early diagnosis of cirrhosis remains challenging, AI-driven models offer a promising solution for early detection, stage estimation, and predicting disease progression. However, well-curated MRI datasets for cirrhosis are still insufficient, which restricts advanced image analysis methods such as radiomics and deep learning, which require large amounts of data and labels~\cite{hantze2024mrsegmentator}. To bridge this gap, a large-scale multi-sequence MRI dataset, \textit{CirrMRI600+}, has been created for liver cirrhosis research~\cite{jha2024cirrmri600+}. This dataset aims to enhance the application of deep learning in analyzing cirrhosis and its stages, improve early-stage diagnostic accuracy in clinical practice, enhance patient outcomes, and encourage further research.


In this study, we address these challenges through a comprehensive deep learning framework specifically designed for multi-sequence MRI analysis. Given an input MRI scan $X$ comprising T1W and T2W sequences, we develop a mapping function $f_\theta$ that estimates the cirrhosis stage $y$ while accounting for the complementary information present across sequences. The framework incorporates sequence-specific feature extraction, cross-sequence attention mechanisms, and a unified classification architecture. The technical contributions of our work advance the field in several key directions: 



\begin{itemize}
\item To our knowledge, this work presents the first comprehensive deep learning framework for automated liver cirrhosis staging using multi-sequence MRI data. While previous studies have explored either radiomics features or limited deep learning approaches for cirrhosis detection, none have systematically addressed the challenging task of accurate stage estimation across a large(r)-scale, multi-center dataset.
\item  We performed an extensive empirical evaluation across T1W and T2W MRI sequences, systematically analyzing a broad spectrum of deep learning architectures ranging from established networks (VGG-19, ResNet variants) to recent state-of-the-art models (MambaVision, ConvNext). Our comprehensive experiments encompassed six distinct deep learning architectures and several radiomics-based machine learning approaches. The results demonstrate the superior capability of deep learning methods, with our best model achieving 72.8\% accuracy on T1W sequences, significantly outperforming traditional radiomics approaches (54.0\%). Notably, we observe that deeper architectures consistently excel at capturing subtle tissue variations across cirrhosis stages, though their performance varies markedly, highlighting the inherent challenges in distinguishing intermediate disease progression.
   \end{itemize}


\section{Methods}
\label{sec:experiments}

\begin{table*}[t!]
    \centering
    \caption{Performance comparison of deep learning models on CirrMRI600+ T1W and T2W MRI dataset.}
    
    \begin{tabular} 
    {c|c|c|c|c|c|c|c}
    \toprule
        
    \textbf{{Dataset}}  & \textbf{{Model}}
    & \textbf{Param(M)} 
    &\textbf{Acc$\uparrow$}  &\textbf{Prec(\%)$\uparrow$} & \textbf{Sens(\%)$\uparrow$} & \textbf{Spec(\%)$\uparrow$}  & \textbf{F1(\%)$\uparrow$}  
    \\ 
    \hline
     
    \multirow{6}{*}{T1W} 
    
    &VGG-19~\cite{simonyan2014very}   
    &139.58 &\textbf{0.728} &\textbf{70.23} &\textbf{68.80} &\textbf{85.86} &\textbf{69.36} \\
    
    &ResNet-50~\cite{he2016deep}  
    &23.51 &0.611 &53.87 &56.79 &80.44 &54.82 \\
    

    
    
    &PVTv2-B2~\cite{wang2022pvt}       
    &24.85 &0.652 &66.41 &62.26 &82.42 &63.13 \\
    
    &Vim-S~\cite{zhu2024vision}  
    &25.44 &0.563 &57.59 &54.05 &77.45 &54.98 \\
    
    &MedMamba-S~\cite{yue2024medmamba} 
    &18.62 &0.490 &48.57 &46.97 &75.01 &47.20 \\
    
    &MambaVision-T~\cite{hatamizadeh2024mambavision}
    &31.16 &0.545 &56.35 &49.49 &76.17 &51.33 \\ \hline

    \multirow{6}{*}{T2W}    
    &VGG-19~\cite{simonyan2014very}   
    &139.58 &0.574 &54.10 &49.10 &79.37 &50.36 \\
    
    &ResNet-50~\cite{he2016deep}  
    &23.51 &0.574 &49.50 &48.93 &78.72 &49.20 \\
    

    
    
    &PVTv2-B2~\cite{wang2022pvt}       
    &24.85 &0.544 &49.09 &49.97 &77.15 &49.10 \\
    
    &Vim-S~\cite{zhu2024vision}  
    &25.44 &0.485 &52.56 &49.74 &75.50 &48.83 \\
    
    &MedMamba-S~\cite{yue2024medmamba} 
    &18.62 &0.506 &53.04 &48.93 &76.58 &49.10 \\
    
    &MambaVision-T~\cite{hatamizadeh2024mambavision}
    &31.16 &\textbf{0.638} &\textbf{59.13} &\textbf{58.01} &\textbf{81.38} &\textbf{58.43} \\
        
\bottomrule
\end{tabular}
\label{tab:deeplearning_t1w}
\end{table*}

\textbf{Dataset}: 
Our framework addresses the challenging task of liver cirrhosis stage estimation through a comprehensive deep learning approach. The CirrMRI600+ dataset forms the foundation of our work, comprising 628 high-resolution abdominal MRI scans from 339 patients diagnosed with liver cirrhosis. This diverse dataset includes 310 T1W and 318 T2W scans, providing rich multi-sequence information for stage classification. We employ an 8:1:1 split ratio, allocating 234 samples for training, 29 for validation, and 28 for testing, ensuring robust evaluation of our methods.  \textbf{Fig.1} presents examples of T1W and T2W MRI scans for mild, moderate, and severe cirrhotic cases.


\begin{figure}[htbp]
\centering
\includegraphics[width=\linewidth]{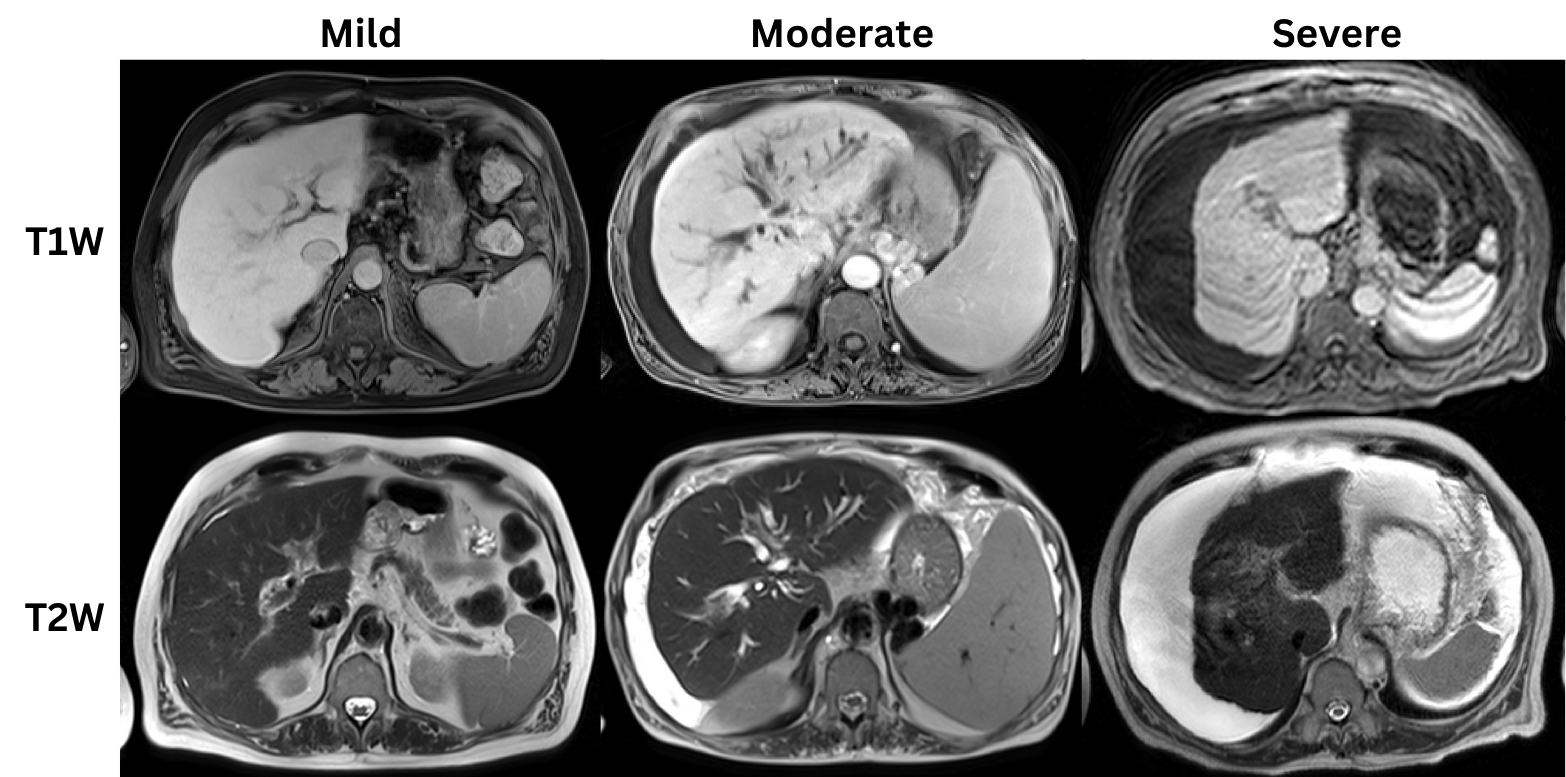}
\caption{T1W and T2W MRI scans of cirrhotic patients are shown for mild, moderate, and severe cases. }
\label{fig:liver}
\end{figure}

\textbf{Network Architecture:} Our architectural design prioritizes two key objectives: effective sequence-specific feature extraction and robust stage classification. We evaluate six state-of-the-art deep learning architectures, including established networks like VGG-19 and ResNet variants, alongside emerging architectures such as \textit{MambaVision} and \textit{ConvNext}. Each architecture processes the multi-sequence MRI inputs through specialized encoding pathways before classification.
For sequence-specific processing, we implement independent encoding streams for T1W and T2W sequences. The feature extraction pathway incorporates residual connections and dense blocks, enabling efficient gradient flow while maintaining fine-grained spatial information critical for stage differentiation. We express this mathematically as:
$F_t = E_t(X_t; \theta_t)$, where $E_t$ represents the encoder for sequence type $t$, $X_t$ is the input sequence, and $\theta_t$ denotes the learnable parameters.

\textbf{Implementation details:} In this work, we implemented deep learning models using the PyTorch~\cite{paszke2019pytorch} framework. All experiments were conducted on an NVIDIA A100 GPU. We employed the AdamW optimizer to optimize model parameters, with the learning rate of 1e$^{-4}$ and the batch size of 32. We train each model for 100 epochs with early stopping patience of 20 epochs to prevent overfitting. The loss function combines cross-entropy for classification with an auxiliary attention guidance term:
$$L = L_{CE}(y, \hat{y}) + \lambda L_{att}(A, A_{gt}),$$
where $L_{CE}$ represents the cross-entropy loss, $L_{att}$ guides attention map learning, $\lambda$ balances their contributions, and $\hat{y}$ is predicted decision while $y$ is ground truth.

\begin{table*}[t!]
\centering
\caption{Performance of deep learning models for different liver cirrhosis stages on CirrMRI600+ T1W MRI dataset.}  
 \begin{tabular} {c|c|c|c|c|c}
\toprule

\textbf{{Liver cirrhosis stage}}  & \textbf{{Model}}  & \textbf{Prec (\%)$\uparrow$}  &\textbf{Sens(\%)$\uparrow$} 
&\textbf{Spec(\%)$\uparrow$} 
& \textbf{F1(\%)$\uparrow$} 
\\
\hline

    \multirow{6}{*}{Mild} 
    &VGG-19~\cite{simonyan2014very}   
    &\textbf{77.83} &\textbf{84.31} &\textbf{78.97} &\textbf{80.94} \\
    
    &ResNet-50~\cite{he2016deep} 
    &73.51 &66.67 &78.97 &69.92 \\
    

    
    &PVTv2-B2~\cite{wang2022pvt} &74.04 &75.49 &76.82 &74.76 \\
    
    &Vim-S~\cite{zhu2024vision} 
    &62.44 &62.75 &66.95 &62.59 \\
    
    &MedMamba-S~\cite{yue2024medmamba} 
    &67.09 &51.96 &77.68 &58.56 \\
    
    &MambaVision-T~\cite{hatamizadeh2024mambavision} 
    &61.37 &70.10 &61.37 &65.45 \\
    
    \hline
    
    \multirow{6}{*}{Moderate} 
    &VGG-19~\cite{simonyan2014very}   
    &\textbf{48.51} &48.04 &\textbf{84.48} &\textbf{48.28} \\
    
    &ResNet-50~\cite{he2016deep} &20.24 &16.67 &80.00 &18.28 \\
    
    
    
    &PVTv2-B2~\cite{wang2022pvt} &37.41 &\textbf{50.98} &74.03 &43.15 \\
    
    &Vim-S~\cite{zhu2024vision} 
    &31.39 &42.16 &71.94 &35.98 \\
    
    &MedMamba-S~\cite{yue2024medmamba} 
    &20.83 &29.41 &65.97 &24.39 \\
    
    &MambaVision-T~\cite{hatamizadeh2024mambavision} 
    &21.60 &26.47 &70.75 &23.79 \\
    
    \hline
    \multirow{6}{*}{Severe} 
    
    &VGG-19~\cite{simonyan2014very}   
    &84.35 &74.05 &94.12 &\textbf{78.86} \\
    
    &ResNet-50~\cite{he2016deep} &67.86 &\textbf{87.02} &82.35 &76.25 \\
    

    
    &PVTv2-B2~\cite{wang2022pvt} &\textbf{87.78} &60.31 &\textbf{96.41} &71.49 \\
    
    &Vim-S~\cite{zhu2024vision} 
    &78.95 &57.25 &93.46 &66.37 \\
    
    &MedMamba-S~\cite{yue2024medmamba} 
    &57.78 &59.54 &81.37 &58.65 \\
    
    &MambaVision-T~\cite{hatamizadeh2024mambavision} 
    &86.08 &51.91 &96.41 &64.76 \\
    
\bottomrule
\end{tabular}
\label{tab:deeplearningforallstages_t1}
\end{table*}

\begin{table*}[t!]
\centering
\caption{Performance of deep learning models for different liver cirrhosis stages on CirrMRI600+ T2W  MRI dataset.}  
 \begin{tabular} {c|c|c|c|c|c}
\toprule

\textbf{{Liver cirrhosis stage}}  & \textbf{{Model}}  & \textbf{Prec (\%)$\uparrow$}  &\textbf{Sens(\%)$\uparrow$} 
&\textbf{Spec(\%)$\uparrow$} 
& \textbf{F1(\%)$\uparrow$} 
\\
\hline

    \multirow{6}{*}{Mild} 
    &VGG-19~\cite{simonyan2014very}   
    &\textbf{82.03} &\textbf{78.68} &81.61 &\textbf{80.32} \\
    
    &ResNet-50~\cite{he2016deep} 
    &78.78 &76.80 &77.93 &77.78 \\
    

    
    
    &PVTv2-B2~\cite{wang2022pvt} &72.73 &60.19 &75.92 &65.87 \\
    
    &Vim-S~\cite{zhu2024vision} 
    &71.65 &43.57 &81.61 &54.19 \\
    
    &MedMamba-S~\cite{yue2024medmamba} 
    &77.33 &54.55 &\textbf{82.94} &63.97 \\
    
    &MambaVision-T~\cite{hatamizadeh2024mambavision} 
    &77.71 &76.49 &76.59 &77.09 \\
    
    \hline
    
    \multirow{6}{*}{Moderate} 
    &VGG-19~\cite{simonyan2014very}   
    &19.32 &30.08 &65.57 &23.53 \\
    
    &ResNet-50~\cite{he2016deep} &17.24 &18.80 &75.26 &17.99 \\
    

    
    
    &PVTv2-B2~\cite{wang2022pvt} &13.16 &15.04 &72.78 &14.04 \\
    
    &Vim-S~\cite{zhu2024vision} 
    &22.05 &\textbf{43.61} &57.73 &29.29 \\
    
    &MedMamba-S~\cite{yue2024medmamba} 
    &22.35 &42.86 &59.18 &29.38 \\
    
    &MambaVision-T~\cite{hatamizadeh2024mambavision} 
    &\textbf{30.77} &36.09 &\textbf{77.73} &\textbf{33.22} \\
    
    \hline
    \multirow{6}{*}{Severe} 
    
    &VGG-19~\cite{simonyan2014very}   
    &60.95 &38.55 &\textbf{90.93} &47.23 \\
    
    &ResNet-50~\cite{he2016deep} &52.47 &51.20 &82.96 &51.83 \\
    
    
    
    
    &PVTv2-B2~\cite{wang2022pvt} &61.39 &\textbf{74.70} &82.74 &\textbf{67.39} \\
    
    &Vim-S~\cite{zhu2024vision} 
    &63.98 &62.05 &87.17 &63.00 \\
    
    &MedMamba-S~\cite{yue2024medmamba} 
    &59.42 &49.40 &87.61 &53.95 \\
    
    &MambaVision-T~\cite{hatamizadeh2024mambavision} 
    &\textbf{68.92} &61.45 &89.82 &64.97 \\
    
\bottomrule
\end{tabular}
\label{tab:deeplearningforallstages}
\end{table*}

\begin{table}[t!]
\centering
\caption{Performance comparison of ResNet and MambaVision variants.}
\begin{tabular} {c|c|c|c}
\toprule
\textbf{Dataset} & \textbf{Model} & \textbf{Pre-trained} & \textbf{Acc$\uparrow$} 
\\ 
\hline
    
    \multirow{12}{*}{CirrMRI600+}  
    
    &ResNet-18    
    &\ding{55} &0.584  \\
    
    &ResNet-18    
    &\ding{51}  &0.557	\\
    
    &ResNet-34             
    &\ding{55}  &0.513	\\
    
    &ResNet-34  
    &\ding{51}  &0.561	\\
    
    &ResNet-50  
    &\ding{55}  &0.545	\\
    
    &ResNet-50  
    &\ding{51}  &0.574	\\
    
    
     &MambaVision-T    
    &\ding{55} &0.638  \\
    
    &MambaVision-T     
    &\ding{51}  &0.610	\\
    
    &MambaVision-T2             
    &\ding{55}  &0.518	\\   
    
    &MambaVision-T2            
    &\ding{51}  &0.500	\\   
    
    &MambaVision-S
    &\ding{55}  &0.500	\\
    
    &MambaVision-S
    &\ding{51}  &0.578	\\

    \bottomrule
\end{tabular}
\label{tab:resnet}
\end{table}


    
    
    
    
    
    
    
        

\begin{table*}[t!]
\centering
\caption{Performance comparison of machine learning models on CirrMRI600+ T1W and T2W MRI datasets using radiomics features.}
 \begin{tabular} {c|c|c|c|c|c|c}
\toprule

\textbf{{Dataset}}  & \textbf{{Model}}  &\textbf{Acc$\uparrow$}   &\textbf{Prec(\%)$\uparrow$} & \textbf{Sens(\%)$\uparrow$} & \textbf{Spec(\%)$\uparrow$}  & \textbf{F1(\%)$\uparrow$}  
\\ 
\hline

    \multirow{6}{*}{T1W}  
    
    &Decision Tree~\cite{quinlan1986induction}    
    &0.391 &42.87 &36.90 &69.50 &37.04 \\
    
    &Random Forest~\cite{breiman2001random}     
    &0.519 &\textbf{57.68} &49.48 &75.35 &49.12 \\
    
    &KNeighbors~\cite{cover1967nearest} 
    &0.478 &50.18 &45.87 &73.16 &46.09 \\
    
    &SVC~\cite{cortes1995support}&0.522 &54.18 &\textbf{50.71} &\textbf{75.99} &\textbf{50.84} \\
    
    &GaussianNB~\cite{rish2001empirical}
    &\textbf{0.540} &47.18 &48.93 &75.52 &47.39 \\
    
    &Logistic Regression~\cite{walker1967estimation}
    &0.508 &53.14 &49.05 &75.27 &49.40 \\ \hline 
    

    \multirow{6}{*}{T2W} 
     
 	&Decision Tree~\cite{quinlan1986induction}    
    &0.380 &38.88 &37.75 &69.29 &36.31 \\
    
    &Random Forest~\cite{breiman2001random}     
    &0.456 &45.62 &43.35 &72.40 &41.88 \\
    
    &KNeighbors~\cite{cover1967nearest} 
    &0.434 &42.38 &40.59 &70.34 &38.06 \\
    
    &SVC~\cite{cortes1995support}      
    &0.510 &\textbf{56.86} &48.46 &75.21 &46.96 \\
    
    &GaussianNB~\cite{rish2001empirical}
    &0.416 &42.60 &43.00 &70.40 &41.10 \\
    
    &Logistic Regression~\cite{walker1967estimation}
    &\textbf{0.528} &55.12 &\textbf{49.50} &\textbf{75.93} &\textbf{48.13} \\

\bottomrule
\end{tabular}
\label{tab:machinelearning_t1}
\end{table*}



    
    
    
    
    
    
    


\begin{table*}[t!]
\centering
\caption{Performance of machine learning models for different liver cirrhosis stages on CirrMRI600+ T1W MRI dataset.}  
 \begin{tabular} {c|c|c|c|c|c}
\toprule

\textbf{{Liver cirrhosis stage}}  & \textbf{{Model}}  & \textbf{Prec (\%)$\uparrow$}  &\textbf{Sens(\%)$\uparrow$} 
&\textbf{Spec(\%)$\uparrow$} 
& \textbf{F1(\%)$\uparrow$} 
\\
\hline

   \multirow{6}{*}{Mild}  
    
  &Decision Tree~\cite{quinlan1986induction}   
 &53.97 &50.00 &62.66 &51.91 \\
    
    &Random Forest~\cite{breiman2001random}     
    &61.29 &65.20 &63.95 &\textbf{63.18} \\
    
    &KNeighbors~\cite{cover1967nearest}
    &56.31 &56.86 &61.37 &56.59 \\
    
    &SVC~\cite{cortes1995support}&\textbf{63.83} &58.82 &\textbf{70.82} &61.22 \\
    
    &GaussianNB~\cite{rish2001empirical}
    &60.36 &\textbf{65.69} &62.23 &62.91 \\
    
    &Logistic Regression~\cite{walker1967estimation}
    &62.63 &58.33 &69.53 &60.41 \\
    

    \hline
    
     \multirow{6}{*}{Moderate} 
    
    &Decision Tree~\cite{quinlan1986induction}    
    &19.47 &36.27 &54.33 &25.34 \\
    
    &Random Forest~\cite{breiman2001random}     
    &\textbf{31.36} &\textbf{51.96} &65.37 &\textbf{39.11} \\
    
   &KNeighbors~\cite{cover1967nearest} 
    &28.48 &44.12 &66.27 &34.62 \\
    
    &SVC~\cite{cortes1995support}&30.49 &49.02 &65.97 &37.59 \\
    
    &GaussianNB~\cite{rish2001empirical}
    &25.42 &14.71 &\textbf{86.87} &18.63 \\
    
    &Logistic Regression~\cite{walker1967estimation}
    &28.48 &46.08 &64.78 &35.21 \\
    
    
    \hline
    \multirow{6}{*}{Severe} 
    
   &Decision Tree~\cite{quinlan1986induction}    
    &55.17 &24.43 &91.50 &33.86 \\
    
    &Random Forest~\cite{breiman2001random}     
    &\textbf{80.39} &31.30 &\textbf{96.73} &45.05 \\
    
    &KNeighbors~\cite{cover1967nearest} 
    &65.75 &36.64 &91.83 &47.06 \\
    
    &SVC~\cite{cortes1995support}&68.24 &44.27 &91.18 &53.70 \\
    
    &GaussianNB~\cite{rish2001empirical}
    &55.77 &\textbf{66.41} &77.45 &\textbf{60.63} \\

    &Logistic Regression~\cite{walker1967estimation}
    &68.29 &42.75 &91.50 &52.58 \\
    

\bottomrule
\end{tabular}
\label{tab:machinelearningforallstages_t1}
\end{table*}

\begin{table*}[t!]
\centering
\caption{Performance of machine learning models for different liver cirrhosis stages on CirrMRI600+ T2W MRI dataset.}  
 \begin{tabular} {c|c|c|c|c|c}
\toprule

\textbf{{Liver cirrhosis stage}}  & \textbf{{Model}}  & \textbf{Prec (\%)$\uparrow$}  &\textbf{Sens(\%)$\uparrow$} 
&\textbf{Spec(\%)$\uparrow$} 
& \textbf{F1(\%)$\uparrow$} 
\\
\hline

   \multirow{6}{*}{Mild}
    
    &Decision Tree~\cite{quinlan1986induction}   
    &58.11 &40.44 &68.90 &47.69 \\
    
    &Random Forest~\cite{breiman2001random}     
    &63.37 &54.23 &66.56 &58.45 \\
    
    &KNeighbors~\cite{cover1967nearest}
    &57.67 &54.23 &57.53 &55.90 \\
    
    &SVC~\cite{cortes1995support}&68.07 &60.82 &\textbf{69.57} &64.24 \\
    
    &GaussianNB~\cite{rish2001empirical}
    &54.63 &38.87 &65.55 &45.42 \\
    
    &Logistic Regression~\cite{walker1967estimation}
    &\textbf{69.26} &\textbf{64.26} &\textbf{69.57} &\textbf{66.67} \\
    

    \hline

    \multirow{6}{*}{Moderate}
    
    &Decision Tree~\cite{quinlan1986induction}    
    &22.30 &45.11 &56.91 &29.85 \\
    
    &Random Forest~\cite{breiman2001random}     
    &26.36 &51.13 &60.82 &34.78 \\
    
    &KNeighbors~\cite{cover1967nearest} 
    &26.85 &51.88 &61.24 &35.38 \\
    
    &SVC~\cite{cortes1995support}&28.36 &\textbf{58.65} &59.38 &38.24 \\
    
    &GaussianNB~\cite{rish2001empirical}
    &27.02 &50.38 &62.68 &35.17 \\
    
    &Logistic Regression~\cite{walker1967estimation}
    &\textbf{29.92} &57.14 &\textbf{63.30} &\textbf{39.28} \\
    
    
    \hline
    
    \multirow{6}{*}{Severe}
    
    &Decision Tree~\cite{quinlan1986induction}    
    &36.22 &27.71 &82.08 &31.40 \\
    
    &Random Forest~\cite{breiman2001random}     
    &47.13 &24.70 &89.82 &32.41 \\
    
    &KNeighbors~\cite{cover1967nearest} 
    &42.62 &15.66 &92.26 &22.91 \\
    
    &SVC~\cite{cortes1995support}&\textbf{74.14} &25.90 &\textbf{96.68} &38.39 \\
    
    &GaussianNB~\cite{rish2001empirical}
    &46.15 &\textbf{39.76} &82.96 &\textbf{42.72} \\
    
    &Logistic Regression~\cite{walker1967estimation}
    &66.18 &27.11 &94.91 &38.46 \\
    

\bottomrule
\end{tabular}
\label{tab:machinelearningforallstages}
\end{table*}

\textbf{Evaluation metrics:} 
We employ a comprehensive set of evaluation metrics including accuracy, precision, sensitivity, specificity, and F1-score to thoroughly assess model performance. Stage-specific analysis provides detailed insights into model behavior across different cirrhosis progression levels. The evaluation protocol ensures fair comparison across architectures and with radiomics-based approaches.

\section{Results}
Our experimental evaluation reveals several key insights into the effectiveness of deep learning approaches for cirrhosis stage estimation. The performance analysis spans both sequence types (T1W and T2W) and encompasses multiple architectural variants (6 state-of-the-art deep learning architectures), providing a comprehensive assessment of automated staging capabilities. Our study is the first in this sense, not only estimating cirrhosis stage estimation with MRI, but also having the most comprehensive one.

\subsection{Model Performance Analysis}
The detailed results are presented in Table~\ref{tab:deeplearning_t1w} for T1W and T2W images, respectively. Among the evaluated architectures, VGG-19~\cite{simonyan2014very} demonstrates superior overall performance on T1W sequences, achieving 72.8\% accuracy and 70.23\% precision. This performance significantly exceeds traditional radiomics-based approaches, which achieve maximum accuracy of 54.0\% using GaussianNB classifiers (Tables 5 and 6). The strong performance of VGG-19 can be attributed to its deep feature hierarchy and extensive receptive field, enabling capture of both local texture patterns and global structural changes characteristic of cirrhosis progression. ResNet variants show consistent but varying performance, with ResNet-50 achieving 61.1\% accuracy on T1W sequences. The relationship between model depth and performance is non-monotonic; ResNet-34 occasionally outperforms deeper variants, suggesting that architectural efficiency may be more crucial than raw capacity for this task. MobileNetV3-Small, despite its lightweight design, maintains competitive performance (59.0\% accuracy), indicating the potential for efficient deployment in resource-constrained clinical settings. 


\subsection{Stage- and Sequence-Specific Analysis}
We assessed the models' capabilities in estimating each specific stage of cirrhosis, and presented them in Tables~\ref{tab:deeplearningforallstages_t1} and \ref{tab:deeplearningforallstages} for T1W and T2W, respectively. Detailed analysis reveals significant variation in model performance across cirrhosis stages. All architectures demonstrate stronger performance in identifying mild (precision 82.03\%) and severe cases (precision 60.95\%) compared to moderate stages (precision 19.32\%). This pattern persists across both sequence types and architectural choices, highlighting an inherent challenge in distinguishing intermediate disease progression states. The performance disparity likely reflects the subtle and heterogeneous nature of tissue changes during the transition from mild to moderate cirrhosis.

The comparative analysis between T1W and T2W sequences reveals important differences in their diagnostic value. T1W sequences consistently yield superior performance across all evaluated models, with average accuracy improvements of 5-10\% compared to T2W sequences. This finding aligns with clinical observations regarding the enhanced visibility of fibrotic changes in T1-weighted imaging. 

\subsection{Impact of network depth, parameter counts, and initial model weights}
To evaluate the impact of network depth and parameter count on liver cirrhosis stage classification, we trained multiple variants of ResNet~\cite{he2016deep} and MambaVision with different complexities. The results, presented in Table~\ref{tab:resnet}, indicating that an increase in parameters does not consistently lead to improved model accuracy. The pretrained ResNet-50, which has fewer parameters than ResNet-152, demonstrated superior performance. Conversely, the larger pretrained ResNet-50 outperformed ResNet-34. These observations indicate that optimal performance in classification models requires a judicious selection of network architecture and parameter count, rather than simply increasing model size. Similarly, we investigated the impact of initial model weights. The from-scratch ResNet-18 demonstrated superior accuracy compared to its pre-trained counterpart. Additionally, only one of five pre-trained MambaVision models exhibited a performance improvement compared to its from-scratch equivalent. These findings suggest that feature representations learned from the large-scale ImageNet dataset do not significantly enhance performance for this task. This is likely due to the domain discrepancy between natural images and liver cirrhosis images, as well as the high similarity among cirrhotic tissues at different stages. 

\subsection{Further details on radiomics}
Radiomics results are presented in Tables 5, 6, and 7, respectively. The features across the three stages of cirrhosis are remarkably similar, making it challenging to effectively segregate these feature regions. Even the GaussianNB machine learning model, which achieved the highest accuracy among the machine learning methods, only reached an accuracy of 0.540. Its performance is significantly lower compared to most deep learning methods, underscoring the potential of deep learning on liver cirrhosis stage estimation.


\section{Discussion and Concluding Remarks}
This work presents the first comprehensive evaluation of deep learning and radiomics approaches for automated liver cirrhosis staging using multi-sequence MRI. Through extensive experimentation, we demonstrate the superior capability of deep learning methods in this challenging task. Our best-performing model achieves 72.8\% accuracy on T1W sequences, establishing a new benchmark for automated cirrhosis staging. The successful application of deep learning to cirrhosis staging opens new possibilities for computer-aided diagnosis and monitoring of liver disease progression. 



Several limitations warrant discussion and point toward future research directions. The primary challenge remains the accurate classification of moderate-stage cirrhosis, where performance consistently lags behind other stages across all evaluated architectures. This limitation likely stems from the inherent ambiguity in tissue characteristics during disease progression and suggests the need for more sophisticated feature learning approaches. 


The current study is on pseudo-3D, while computationally efficient, it may miss important volumetric information when slice thickness is large. Hence, future work will explore fully 3D architectures. Additionally, incorporating clinical metadata and longitudinal imaging data could provide valuable context for more accurate staging predictions. Last, but not least, leveraging multiple imaging modalities and clinical reports, large-scale foundation models (FMs) tailored for medical applications hold significant promise for advancing liver cirrhosis stage estimation. However, the data is still limited at the moment and FMs are at suboptimal for the problem tackled in this paper.

\section{Compliance with ethical standards}
This study was performed in line with the principles of the Declaration of Helsinki. Approval was granted by Northwestern University (No. STU00214545).





\begin{thebibliography}{99}


\bibitem{wang2023remodeling} Yeying Wang, Yang Liu, Yi Liu, Jie Zhong, Jing Wang, Lei Sun,
Lei Yu, Yiting Wang, Qinghua Li, Weilin Jin, et al. Remodeling liver microenvironment by l-arginine loaded hollow polydopamine nanoparticles for liver cirrhosis treatment. \textit{Biomaterials}, 295:122028, 2023

\bibitem{huang2023global} Huang, Daniel Q et al. Global epidemiology of cirrhosis—aetiology, trends and predictions. \textit{Nature reviews Gastroenterology \& hepatology}, volume 20, number 6, pages 388--398, 2023

\bibitem{liu2019wjg} Liu, Rong et al. WJG. \textit{World J Gastroenterol}, volume 25, number 12, pages 1432--1530, 2019

\bibitem{lee2024designing} Lee, Brian P et al. Designing clinical trials to address alcohol use and alcohol-associated liver disease: an expert panel Consensus Statement. \textit{Nature Reviews Gastroenterology \& Hepatology}, pages 1--20, 2024

\bibitem{daher2022proportion} Daher, Darine et al. Proportion time covered by hepatocellular carcinoma surveillance in patients with cirrhosis. \textit{Official journal of the American College of Gastroenterology| ACG}, pages 10--14309, 2022

\bibitem{sepanlou2020global} Sepanlou, Sadaf G et al. PThe global, regional, and national burden of cirrhosis by cause in 195 countries and territories, 1990--2017: a systematic analysis for the Global Burden of Disease Study 2017. \textit{The Lancet gastroenterology \& hepatology}, volume 5, number 3, pages 245--266, 2020

\bibitem{lee2023evolution} Lee, Seohyuk and Saffo, Saad. Evolution of care in cirrhosis: preventing hepatic decompensation through pharmacotherapy. \textit{World Journal of Gastroenterology}, volume 29, number 1, pages 61, 2023

\bibitem{sung2021global} Sung, Hyuna et al. Global cancer statistics 2020: GLOBOCAN estimates of incidence and mortality worldwide for 36 cancers in 185 countries. \textit{CA: a cancer journal for clinicians}, volume 71, number 3, pages 209--249, 2021

\bibitem{yang2019global} Yang, Ju Dong et al. A global view of hepatocellular carcinoma: trends, risk, prevention and management. \textit{Nature reviews Gastroenterology \& hepatology}, volume 16, number 10, pages 589--604, 2019

\bibitem{fattovich2004hepatocellular} Fattovich, Giovanna and Stroffolini, Tommaso and Zagni, Irene and Donato, Francesco. Hepatocellular carcinoma in cirrhosis: incidence and risk factors. \textit{World Journal of Gastroenterology}, volume 127, number 5, pages S35--S50, 2004

\bibitem{ullah2023biological} Muhammad Ikram Ullah et al. Biological role of zinc in liver cirrhosis:
an updated review. \textit{Biomedicines}, 11(4):1094, 2023.

\bibitem{premkumar2022overview} Madhumita Premkumar and Anil C Anand. Overview of complications in cirrhosis. \textit{Journal of Clinical and Experimental Hepatology}, 12(4):1150–1174, 2022.

\bibitem{hantze2024mrsegmentator} Hartmut H{\"a}ntze et al. Mrsegmentator: Robust multi-modality segmentation of 40 classes in mri and ct sequences. \textit{arXiv preprint arXiv:2405.06463}, 2024.


\bibitem{jha2024cirrmri600+} Debesh Jha, Onkar Kishor Susladkar, Vandan Gorade, Elif Keles, Matthew Antalek, Deniz Seyithanoglu, Timurhan Cebeci, Halil Ertugrul Aktas, Gulbiz Dagoglu Kartal, Sabahattin Kaymakoglu, et al. Cirrmri600+: Large scale mri collection and segmentation of cirrhotic
liver. \textit{arXiv preprint arXiv:2410.16296}, 2024.

\bibitem{simonyan2014very} Karen Simonyan and Andrew Zisserman. Very deep convolutional networks for large-scale image recognition. \textit{arXiv preprint arXiv:1409.1556}, 2014.


\bibitem{he2016deep} Kaiming He, Xiangyu Zhang, Shaoqing Ren, and Jian Sun. Deep residual learning for image recognition. In \textit{Proceedings of the IEEE Conference on computer vision and pattern recognition (CVPR)}, pages 770–778, 2016.

\bibitem{howard2019searching} Andrew Howard, Mark Sandler, Grace Chu, Liang-Chieh Chen, Bo Chen, Mingxing Tan, Weijun Wang, Yukun Zhu, Ruoming Pang, Vijay Vasudevan, et al. 
Searching for mobilenetv3. In \textit{Proceedings of the IEEE/CVF international conference on computer vision (ICCV)}, pages 1314–1324, 2019.

\bibitem{liu2022convnet} Zhuang Liu, Hanzi Mao, Chao-Yuan Wu, Christoph Feichtenhofer, Trevor Darrell, and Saining Xie. 
A convnet for the 2020s. In \textit{Proceedings of the IEEE/CVF Conference on computer vision and pattern recognition (CVPR)}, pages 11976–11986, 2022.

\bibitem{wang2022pvt} Wenhai Wang, Enze Xie, Xiang Li, Deng-Ping Fan, Kaitao Song, Ding Liang, Tong Lu, Ping Luo, and Ling Shao. PVT v2: Improved baselines with Pyramid Vision Transformer. \textit{Computational Visual Media}, 8(3):415–424, 2022.


\bibitem{zhu2024vision} Lianghui Zhu, Bencheng Liao, Qian Zhang, Xinlong Wang, Wenyu Liu, and Xinggang Wang. Vision mamba: Efficient visual representation learning with bidirectional state space model. \textit{arXiv preprint
arXiv:2401.09417}, 2024.

\bibitem{yue2024medmamba} Yubiao Yue and Zhenzhang Li. Medmamba: Vision mamba for medical image classification. \textit{arXiv preprint arXiv:2403.03849}, 2024.


\bibitem{hatamizadeh2024mambavision} Ali Hatamizadeh and Jan Kautz. Mambavision: A hybrid mamba-transformer vision backbone. \textit{arXiv preprint arXiv:2407.08083}, 2024.

\bibitem{paszke2019pytorch} Adam Paszke, Sam Gross, Francisco Massa, Adam Lerer, James Bradbury, Gregory Chanan, Trevor Killeen, Zeming Lin, Natalia Gimelshein, Luca Antiga, et al. Pytorch: An imperative style, high-performance deep learning library. \textit{Advances in Neural Information
Processing Systems}, 32, 2019.


\bibitem{quinlan1986induction} J. Ross Quinlan. Induction of decision trees. \textit{Machine learning}, 1:81–106, 1986.

\bibitem{breiman2001random} Leo Breiman. Random forests. \textit{Machine learning}, 45:5–32,
2001.

\bibitem{cover1967nearest} Thomas Cover and Peter Hart. Nearest neighbor pattern classification. \textit{IEEE transactions on information theory}, 13(1):21–27, 1967.

\bibitem{cortes1995support} Corinna Cortes. Support-vector networks. \textit{Machine Learning}, 1995.


\bibitem{rish2001empirical} Irina Rish et al. An empirical study of the naive bayes classifier. In \textit{Proceeding of the IJCAI 2001 workshop on empirical methods in artificial intelligence}, volume 3, page 41–46, 2001.


\bibitem{walker1967estimation} Strother H Walker and David B Duncan. Estimation of the probability of an event as a function of several independent variables. \textit{Biometrika},
54(1-2):167–179, 1967.

\bibitem{friedman2001greedy} Jerome H Friedman. Greedy function approximation: a gradient boosting machine. \textit{Annals of Statistics}. 29(5): 1189–1232, 2001.

\end{thebibliography}
\end{document}